\documentclass[twocolumn,showpacs,preprintnumbers,amsmath,amssymb]{revtex4}

\usepackage{graphicx}
\usepackage{dcolumn}
\usepackage{bm}

\begin{document}

\preprint{APS/123-QED}

\title{Measurement of the mechanical loss 
of a cooled reflective coating 
for gravitational wave detection}

\author{Kazuhiro Yamamoto}
 \email{yamak@icrr.u-tokyo.ac.jp}
\author{Shinji Miyoki}
\author{Takashi Uchiyama}
\author{Hideki Ishitsuka}
\author{Masatake Ohashi}
\author{Kazuaki Kuroda}
\affiliation{%
Institute for Cosmic Ray Research, the University of Tokyo,
5-1-5 Kashiwa-no-Ha, Kashiwa, Chiba 277-8582, Japan}%

\author{Takayuki Tomaru}
\author{Nobuaki Sato}
\author{Toshikazu Suzuki}
\author{Tomiyoshi Haruyama}
\author{Akira Yamamoto}
\author{Takakazu Shintomi}
\affiliation{High Energy Accelerator Research Organization, 
1-1 Oho, Tsukuba, Ibaraki 305-0801, Japan}%

\author{Kenji Numata}
\affiliation{%
Department of Physics, the University of Tokyo, 
7-3-1 Hongo, Bunkyo-ku, Tokyo 113-0033, Japan}%

\author{Koichi Waseda}
\affiliation{%
National Astronomical Observatory of Japan,
2-21-1 Osawa, Mitaka, Tokyo 181-8588, Japan}%

\author{Kazuhiko Ito}
\author{Koji Watanabe}
\affiliation{%
Japan Aviation Electronics Industry, Ltd., 
3-1-1 Musashino, Akishima, Tokyo 196-8555, Japan}%

\date{\today}

\begin{abstract}
We have measured the mechanical loss of a dielectric multilayer 
reflective coating 
(ion-beam sputtered SiO$_2$ and Ta$_2$O$_5$) in cooled mirrors. 
The loss was nearly independent of the temperature (4 K $\sim$ 300 K), 
frequency, optical loss, and 
stress caused by the coating, and the details of the manufacturing processes. 
The loss angle was $(4 \sim 6) \times 10^{-4}$.
The temperature independence of this loss implies 
that the amplitude of the coating thermal noise, 
which is a severe limit in any precise measurement, 
is proportional to the square root of the temperature. 
Sapphire mirrors at 20 K 
satisfy the requirement concerning the thermal noise 
of even future interferometric gravitational 
wave detector projects on the ground, for example, LCGT. 
\end{abstract}

\pacs{04.80.Nn; 05.40.Jc; 06.30.Ft; 07.20.Mc; 62.40.+i; 68.35.Gy; 
68.60.Bs; 77.55.+f; 95.55.Ym}

\maketitle

\section{Introduction}

The development and observation of 
several interferometric gravitational wave detectors 
(LIGO \cite{LIGO}, VIRGO \cite{VIRGO}, GEO \cite{GEO}, 
TAMA \cite{TAMA}) on the ground are presently in progress. 
The sensitivity in the observation band of these detectors 
is expected to be limited by the 
thermal noise of the internal modes of the mirrors 
(this thermal noise is also 
a serious problem in laser frequency stabilization 
using a rigid cavity \cite{Numata-freqstab}). 
Since the thermal noise of 
mirrors with less mechanical loss is smaller, 
research was done in an effort to reduce the loss in the mirrors. 
In order to decrease the thermal noise more effectively, 
it was proposed to cool the mirrors \cite{Uchiyama1}. In the Japanese 
future LCGT project \cite{LCGT}, the mirrors will be cooled. 
In Europe, another future cryogenic interferometer project 
is being considered \cite{EGO}. 

For estimating the temperature dependence of the thermal noise to evaluate 
the adequate mirror temperature for future cryogenic projects, 
measuring the loss in the low-temperature region is necessary.
The loss of cooled sapphire (above 4 K), which is the mirror substrate of 
LCGT, has already been measured \cite{Uchiyama2}. 
Recent theoretical  
\cite{Levin,Harry,Yamamoto-Amaldi,Nakagawa,Yamamoto2} 
and experimental \cite{Numata3,Yamamoto3,Black} work
has revealed that the loss of the reflective coating on 
the mirror surface has a large contribution to the thermal noise. 
The loss of a cooled coating is also an interesting issue 
in solid state physics (for example, Ref. \cite{Vu}). 
Nevertheless, a low-temperature measurement of 
the mechanical loss in the mirror reflective coating for 
gravitational wave detection 
had never been reported before ours. 

We measured the mechanical loss of a cooled coating. 
The measured loss was 
almost constant between 4 K and 300 K. Thus, the amplitude of the coating 
thermal noise is proportional to the square root of the temperature. 
At 20 K, the summation of the thermal noise of the coating and 
sapphire substrate loss is sufficiently smaller 
than the goal sensitivity 
of LCGT.
Our measurement provides some clues 
about the properties of the coating material at low temperature. 

\section{Experimental method}

\subsection{Outline}

In order to evaluate the mechanical loss of the coating, 
we prepared sapphire disks 
with and without a coating. 
After measuring the decay time of 
the excited resonant motions of these 
disks (ring down method) at low temperature, 
and calculating each Q-value, 
we obtained the loss of the coating 
by comparing these Q-values. 
In this experiment, it was possible to accurately estimate 
the cooled coating loss 
because the loss of the sapphire was extremely 
small at low temperature, 
and a coating was made on the thin disks 
in order to enhance the effect of the coating loss.  

\subsection{Samples}

\begin{figure}
\begin{minipage}{8.6cm}
\includegraphics[width=8.6cm]{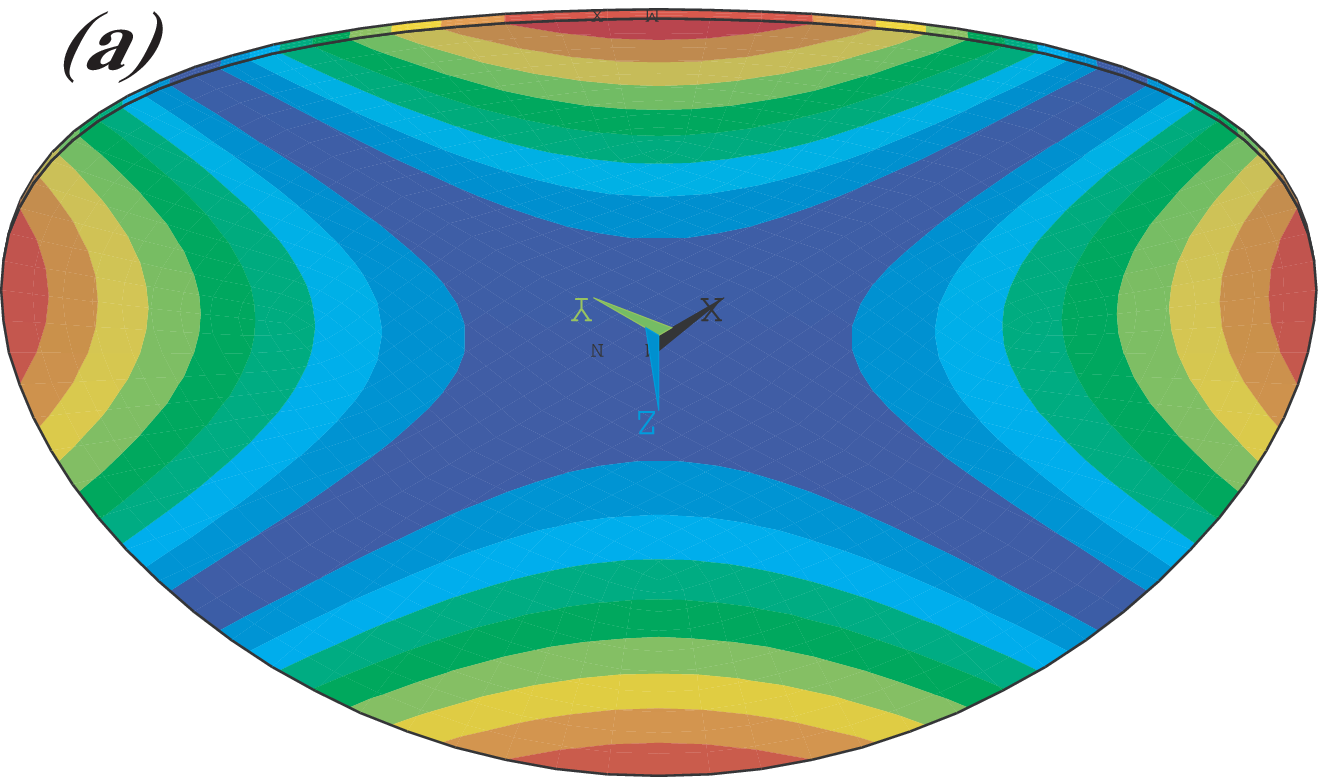}
\end{minipage}
\quad
\begin{minipage}{8.6cm}
\includegraphics[width=8.6cm]{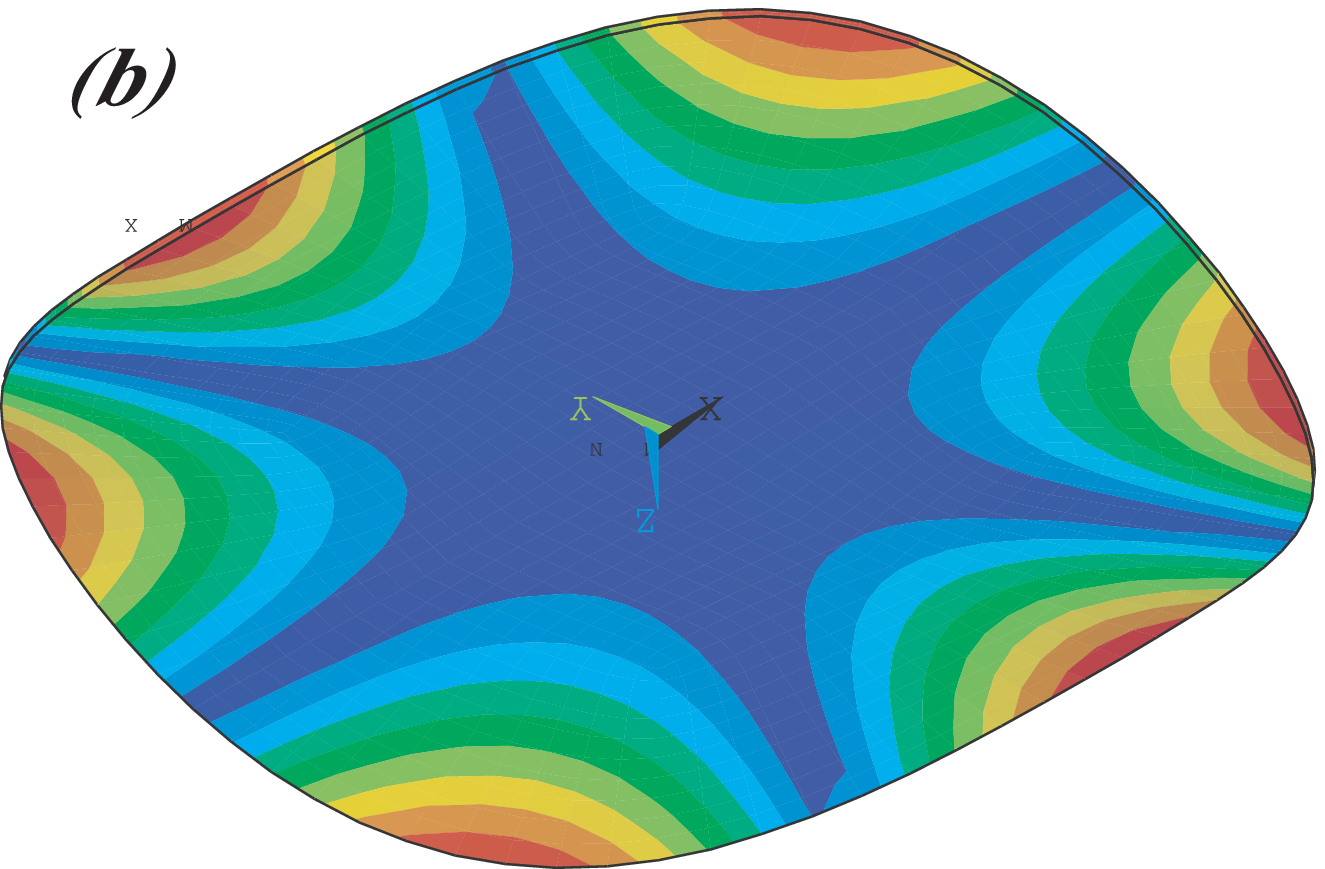}
\end{minipage}
\caption{\label{modeshape}Shapes of the first (a) and third (b) modes.}
\end{figure}

The sapphire disks were supplied by SHINKOSHA \cite{Shinkosya}, 
Japan.
The diameter of these disks was 100 mm. 
In order to observe the disk thickness dependence of the loss, 
two kinds of the disks, 0.5 and 1 mm thick disks, 
were prepared. 
The c-axis was perpendicular to the flat surface. 
Both sides were commercially polished (not super polished).  
The root mean square of the micro-roughness of the surfaces was about 0.1 nm.
The coatings were made on some sapphire disks. 
The other uncoated disks were used to measure the Q-values without the coating.
The specifications of the coating on the sapphire disks were almost the same 
as those of typical mirrors of 
the gravitational wave detectors, as follows. 
These disks were coated by means of ion-beam sputtering. 
This dielectric multilayer reflective coating 
consisted of 31 alternating layers of SiO$_2$ and Ta$_2$O$_5$. 
The total thickness was 4.8 $\mu$m.
The optical thickness of a layer was a quarter of the wavelength, 
which was 1.064 $\mu$m.
The resulting power reflectance was estimated to be 99.99\%.
For investigating the effect of the details in the manufacturing process, 
the coating was made by two venders: 
the National Astronomical Observatory of Japan (NAOJ) 
and Japan Aviation Electronics Industry, Ltd. (JAE). The latter 
made the coating on the mirrors of TAMA \cite{TAMA}. 
The JAE coating \cite{Sato} was superior 
regarding low optical loss compared to that of NAOJ. 
Although the mirrors are usually annealed after the coating, 
only a disk with the JAE coating was annealed in this experiment. 
The specifications of the disks 
are summarized in Table \ref{sample}. 
We measured the Q-values of the first and third modes. 
The resonant frequencies are given in Table \ref{frequency}. 
These frequencies with the coating 
were the same as those without the coating. 
The shapes of the first and third modes are shown in (a) and (b) 
of Fig. \ref{modeshape}, respectively.
\begin{table}
\caption{\label{sample} Specifications of sapphire disks.}
\begin{ruledtabular}
\begin{tabular}{cccc}
&  disk thickness & heat process & coating vender\\
\hline 
Sample 1 & 0.5 mm & not annealed & NAOJ \\
Sample 2 & 1 mm & not annealed & NAOJ \\

Sample 3 & 1 mm & not annealed & JAE \\
Sample 4 & 1 mm & annealed & JAE \\
Sample 5 & 0.5mm & not annealed & not coated \\
Sample 6 & 1 mm & not annealed & not coated \\
\end{tabular}
\end{ruledtabular}
\end{table}

\begin{table}
\caption{\label{frequency}Resonant frequencies of the sapphire disks.}
\begin{ruledtabular}
\begin{tabular}{ccc}
thickness & first mode & third mode \\
\hline 
0.5 mm & 0.52 kHz  & 1.2 kHz \\
1 mm & 1.1 kHz  & 2.5 kHz \\
\end{tabular}
\end{ruledtabular}
\end{table}

\subsection{Measurement apparatus}

\begin{figure}
\includegraphics[width=8.6cm]{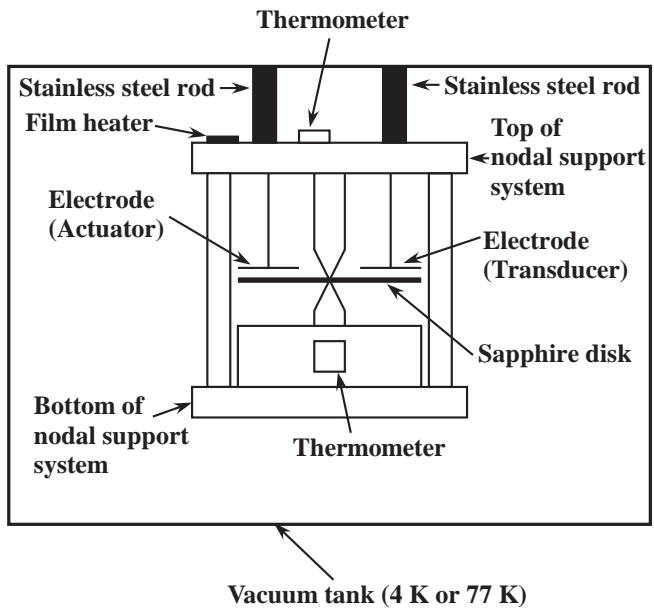}
\caption{\label{nodal} Schematic side view of the measurement apparatus.
Only the center of the sapphire disk was grasped by 
the nodal support system \cite{Numata}. 
This support was made of copper. 
We used an electrostatic 
actuator to excite the resonant vibration. 
The ring down of this resonant motion was monitored 
by an electrostatic transducer. 
The top of the nodal support system was connected by 
stainless-steel rods to a vacuum tank. 
This chamber was immersed 
into liquid helium or nitrogen. 
The pressure in the cooled chamber was between $10^{-5}$ Pa and $10^{-3}$ Pa
in most cases. 
A film heater was put on the top of the nodal support system 
to control the temperature between 4 K and 30 K. 
The two thermometers were fixed at the top and the bottom of 
the nodal support system.}
\end{figure}
Figure \ref{nodal} is a schematic side view of the measurement apparatus.
We adopted a nodal support system \cite{Numata} to grasp the sapphire disk. 
In this system, only the center of the disk was fixed 
(the diameter of the contact area 
between the sapphire disk and the support was 2 mm). 
Since the center is the nodal point in almost all 
resonant modes, 
the contamination 
of the loss of the support system, itself, was small. 
All parts of this support were made of copper, which has a high thermal 
conductivity at cryogenic temperatures. 
We used an electrostatic 
actuator to excite the resonant vibration. 
The ring down of this resonant motion was monitored 
by an electrostatic transducer \cite{transducer}. 
The bias voltage of the actuator and the transducer was 
a few hundreds volts. 
The loss caused by this transducer was negligible 
because the measured Q-values were independent of the bias voltage. 

The top of the nodal support system was connected by 
stainless-steel rods to a vacuum tank. 
This chamber was immersed 
into liquid helium or nitrogen. 
The pressure in the cooled chamber was between $10^{-5}$ Pa and $10^{-3}$ Pa
in most cases. 
In this pressure region, the gas damping was sufficiently small
because the measured Q-values with the coating did not depend on the pressure.
A film heater was put on the top of the nodal support system 
to control the temperature between 4 K and 30 K. 
The two thermometers were fixed at the top and the bottom of 
the nodal support system instead of the sapphire disk because 
the thermometer possibly increases the mechanical loss.
Although, in principle, the temperature of the sapphire disk 
was the same as that
of these thermometers, we confirmed this temperature homogeneity using a 
dummy sapphire disk with the thermometer. 
This dummy disk has never been adopted to measure the Q-values.

\begin{figure}
\begin{minipage}{8.6cm}
\includegraphics[width=8.6cm]{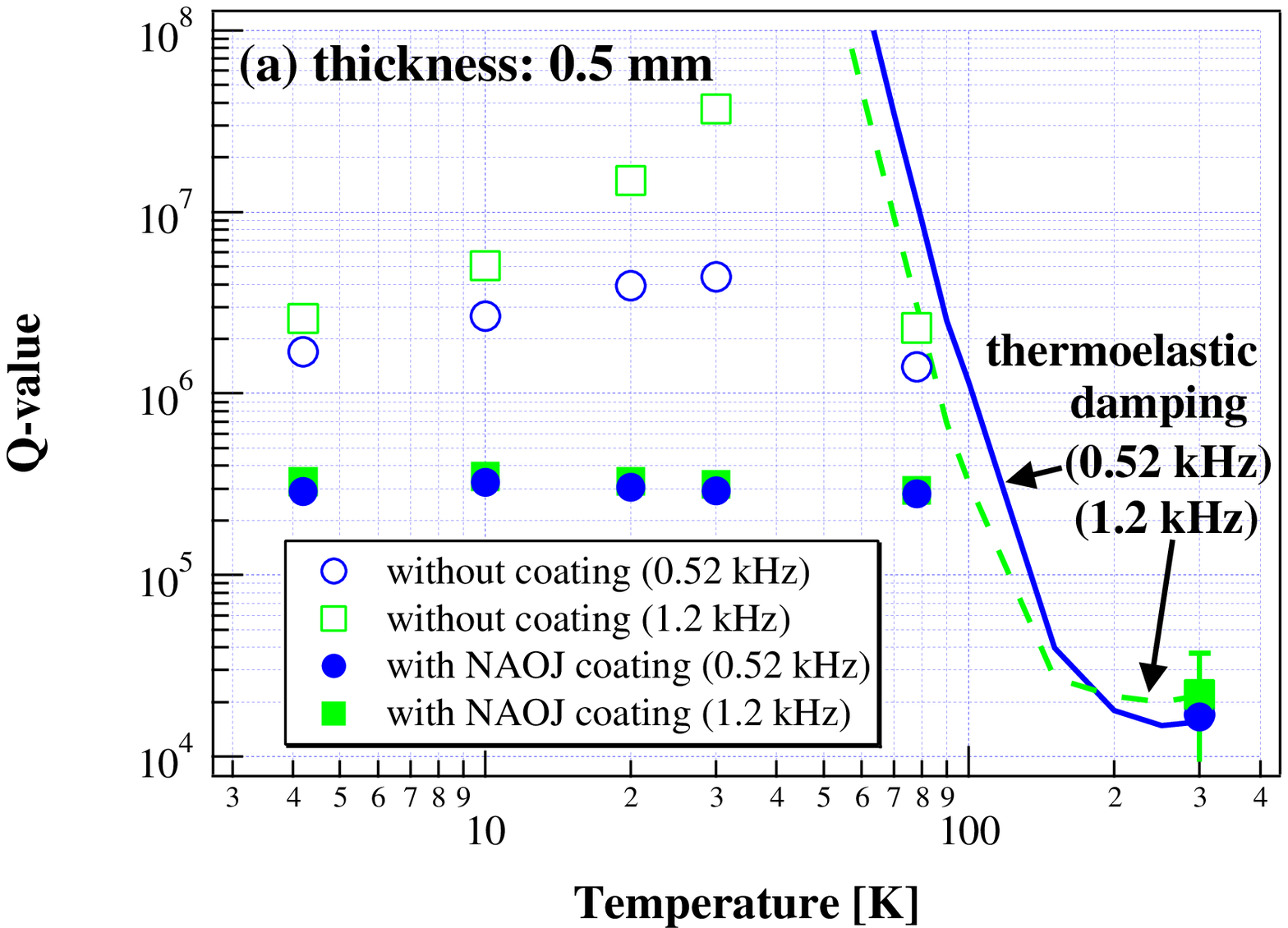}
\end{minipage} 
\quad
\begin{minipage}{8.6cm}
\includegraphics[width=8.6cm]{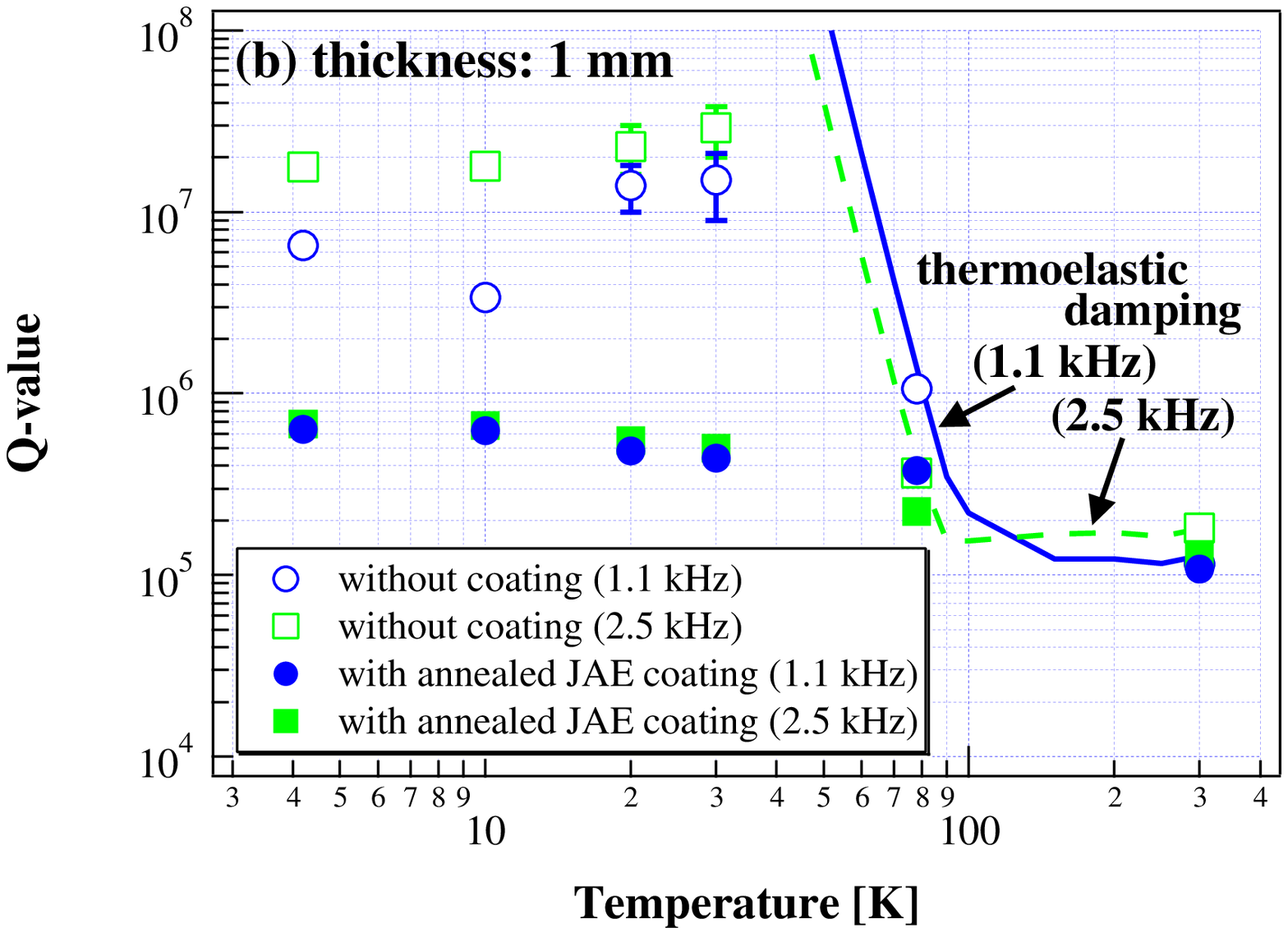}
\end{minipage}
\caption{\label{Q}Typical measured Q-values of the sapphire disks with 
(close marks) and without (open marks) the coating. 
The thicknesses of the disks in graphs (a) and (b) were 0.5 mm
and 1 mm, respectively. 
The coatings in graphs (a) and (b) were NAOJ (Sample 1) 
and annealed JAE (Sample 4).
The disk without the coating in the graphs (a) and (b) 
was Sample 5 and Sample 6.
The circles (blue in online) and squares (green in online) 
are the Q-values of the first and third modes.   
The solid and dashed lines represent the Q-values limited by 
the thermoelastic damping \cite{Zener} of the first and third modes, 
respectively \cite{Blair}.}
\end{figure}
In order to check the ability of our measurement system, 
the typical differences between the measured Q-values 
with (close marks in Fig. \ref{Q}) and without 
(open marks in Fig. \ref{Q}) the coating are introduced.
The thicknesses of the disks in the graphs (a) and (b) of Fig. \ref{Q} 
were 0.5 mm and 1 mm, respectively. 
The coating in the graphs (a) and (b) was NAOJ (Sample 1) 
and annealed JAE (Sample 4).
The disk without the coating in the graphs (a) and (b) 
was Sample 5 and Sample 6.
The circles (blue in online) and squares (green in online) 
are the Q-values of the first and third modes, 
respectively.   
In the low-temperature region, there was a large discrepancy between 
the Q-values with and without the coating \cite{Support loss}.   
An accurate evaluation of the cooled coating loss was possible.  
At room temperature,  
the differences between the Q-values with and without the coating 
were small (the marks with the coating in Fig. \ref{Q} completely overlap 
those without the coating 
except for the third mode of the 1 mm thickness disk). 
The evaluated coating loss angles at 300 K were not 
precise. 
The reason why the differences were small 
was that the thermoelastic damping \cite{Zener} in 
the sapphire disks was large at 300 K. 
The solid and dashed lines in Fig. \ref{Q} represent the Q-values limited by 
the thermoelastic damping of the first and third modes, 
respectively \cite{Blair}. 


\section{Results}


The formula \cite{Harry,Crooks,Penn,Crooks-Amaldi}
of the loss angle, $\phi_{\rm coating}$, which 
is the magnitude of the loss in the coating, derived from the 
measured Q-values  
is described as \cite{Landau,formula}
\begin{equation}
\phi_{\rm coating}=\frac1{3}\frac{d_{\rm disk}}{d_{\rm coating}}
\frac{Y_{\rm disk}}{Y_{\rm coating}}
\left(\frac1{Q_{\rm with}}-\frac1{Q_{\rm without}}\right). 
\label{coating phi}
\end{equation}
The values, $Q_{\rm with}$ and $Q_{\rm without}$, are 
the measured Q-values with and without the coating. 
The parameters 
($d_{\rm disk}$, $d_{\rm coating}$, $Y_{\rm disk}$ and $Y_{\rm coating}$) 
are the thicknesses and 
Young's moduli of the sapphire disk and the coating, respectively. 
These values are summarized in Table \ref{thickness}. 
The Young's modulus of the coating is the average 
of those of SiO$_2$ ($7.2 \times 10^{10}\ $Pa \cite{Sapieha}) 
and Ta$_2$O$_5$ 
($1.4 \times 10^{11}\ $Pa \cite{Sapieha,Martin,CSIRO}) 
\cite{Harry, Crooks, Young}.
\begin{table}
\caption{\label{thickness}Thicknesses and Young's moduli 
of the disks and coating.}
\begin{ruledtabular}

\begin{tabular}{ccc}
& thickness & Young's modulus \\
\hline
sapphire disk & 1 mm or 0.5 mm & $4.0 \times 10^{11}$ Pa \\
coating & 4.8 $\mu$m & $1.1 \times 10^{11}$ Pa \\
\end{tabular}
\end{ruledtabular}
\end{table}

\begin{figure*}
\begin{minipage}{8.6cm}
\includegraphics[width=8.6cm]{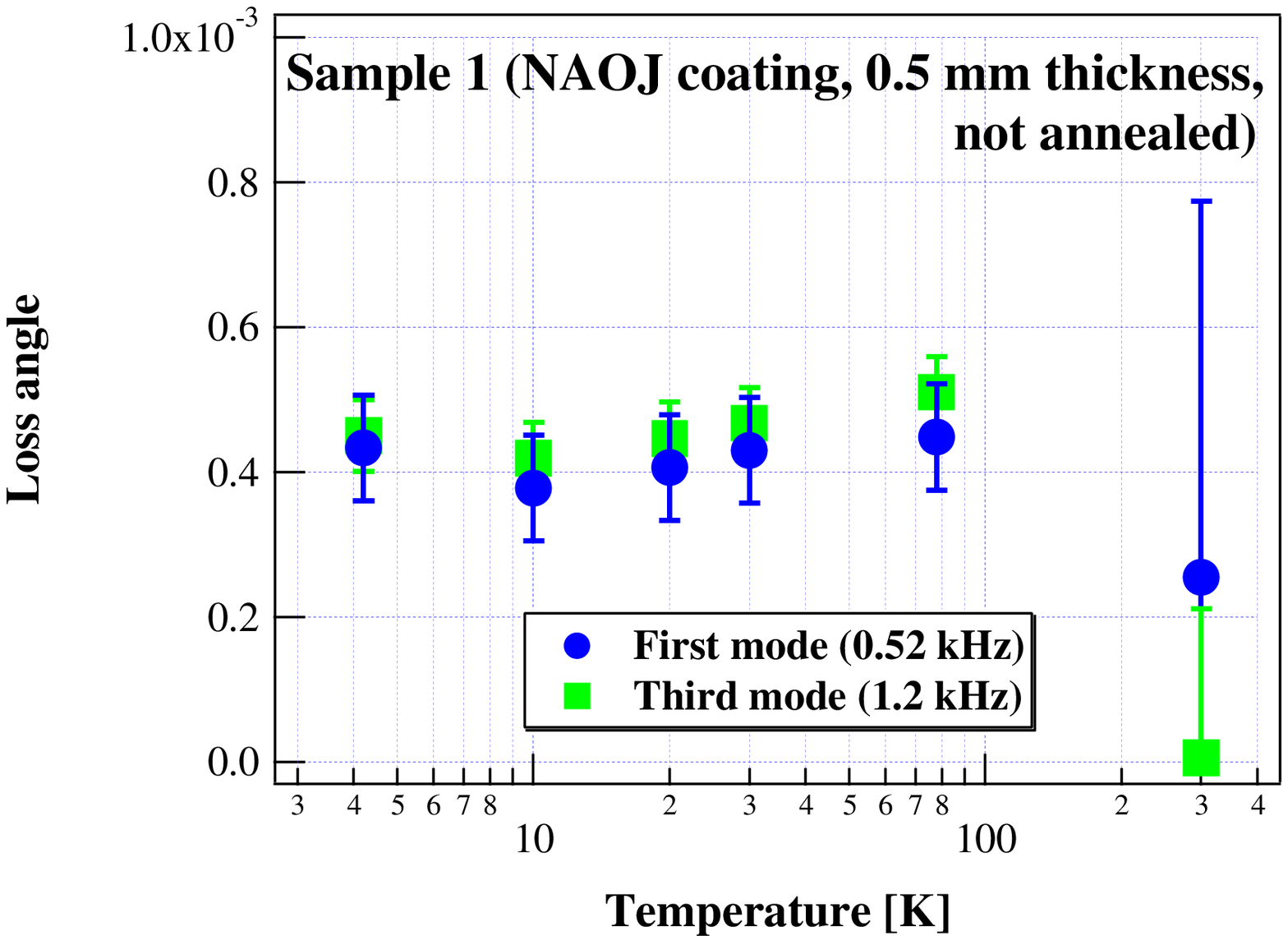}
\end{minipage}
\quad
\begin{minipage}{8.6cm}
\includegraphics[width=8.6cm]{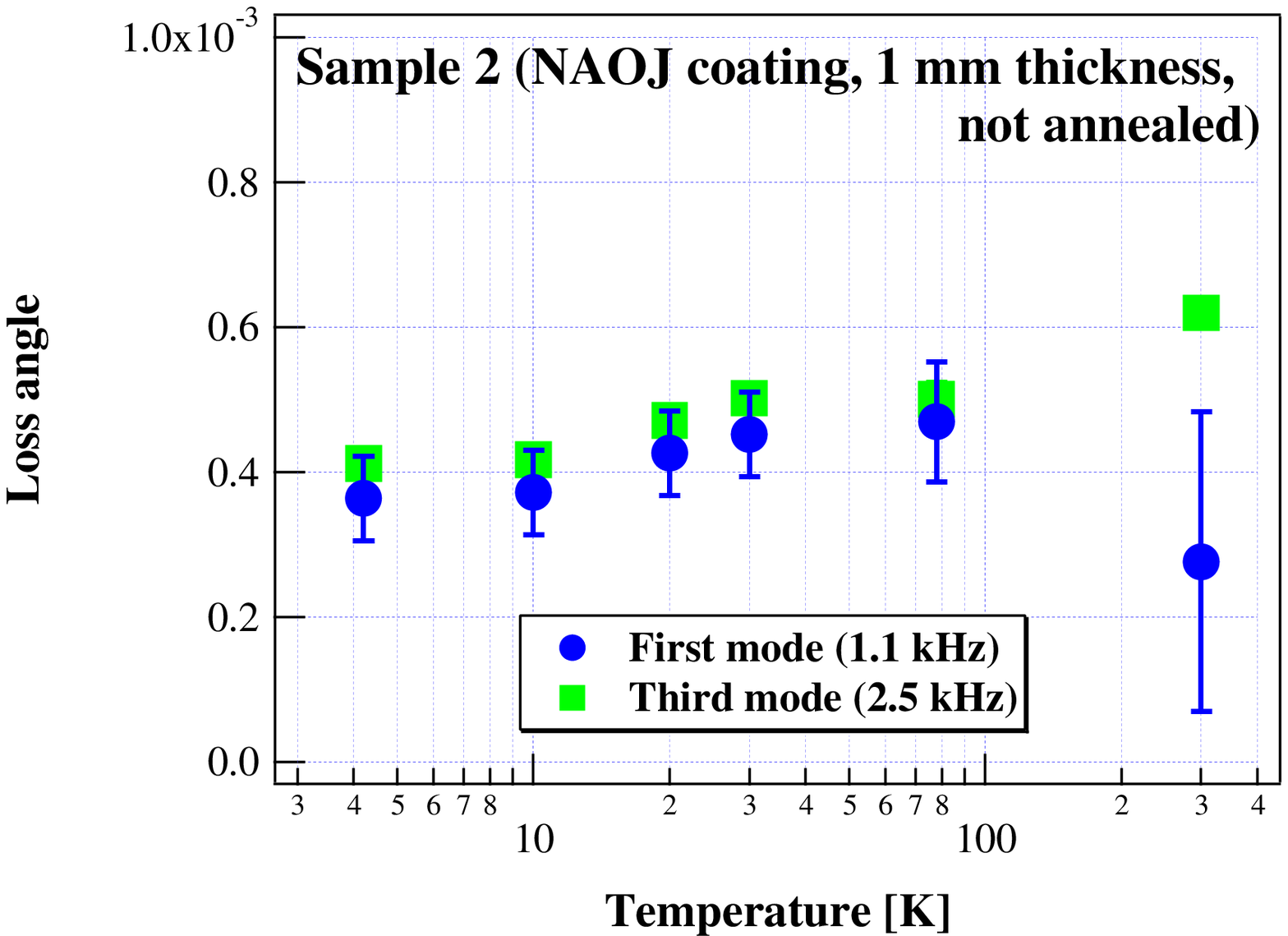}
\end{minipage}
\quad
\begin{minipage}{8.6cm}
\includegraphics[width=8.6cm]{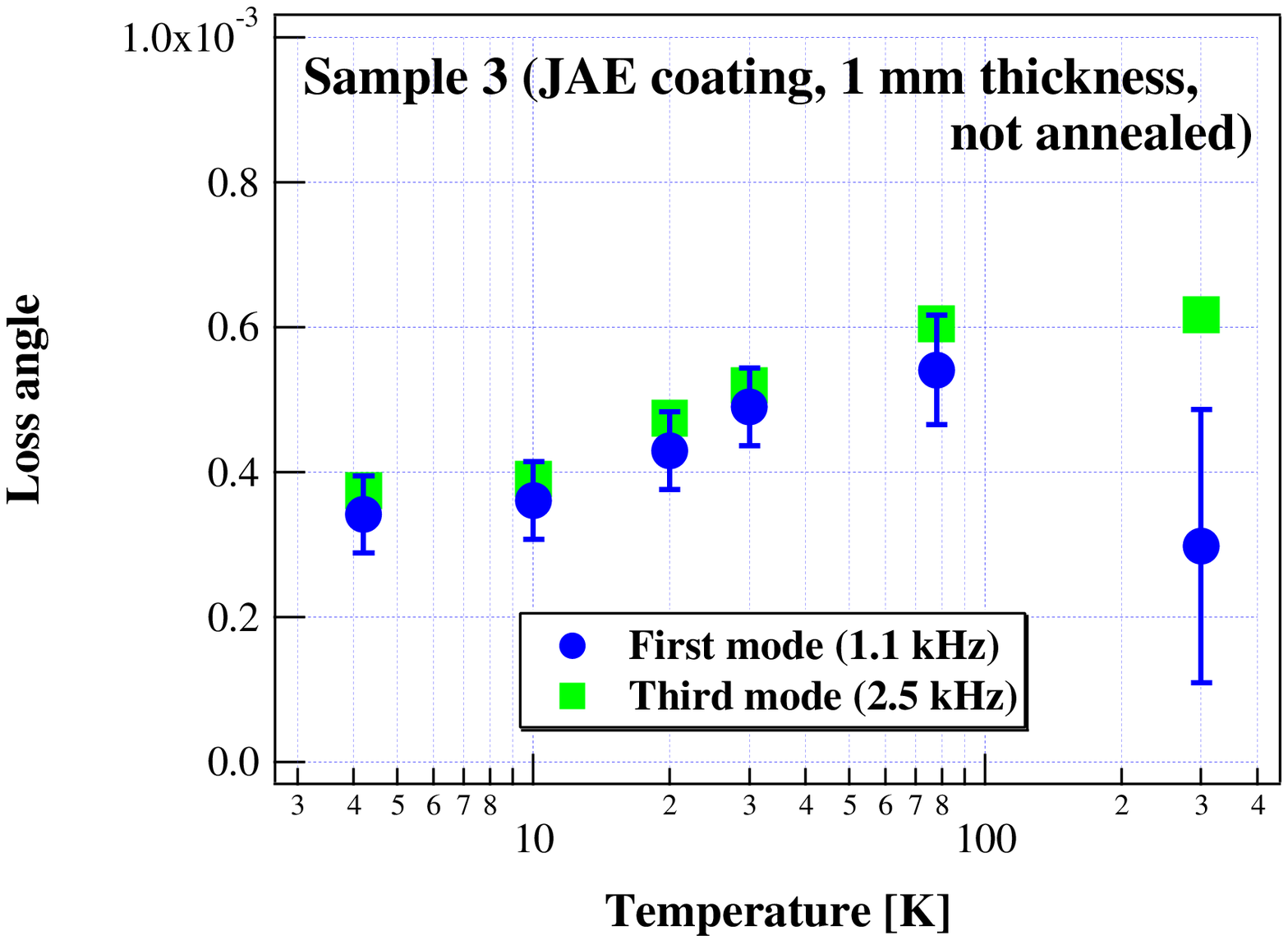}
\end{minipage}
\quad
\begin{minipage}{8.6cm}
\includegraphics[width=8.6cm]{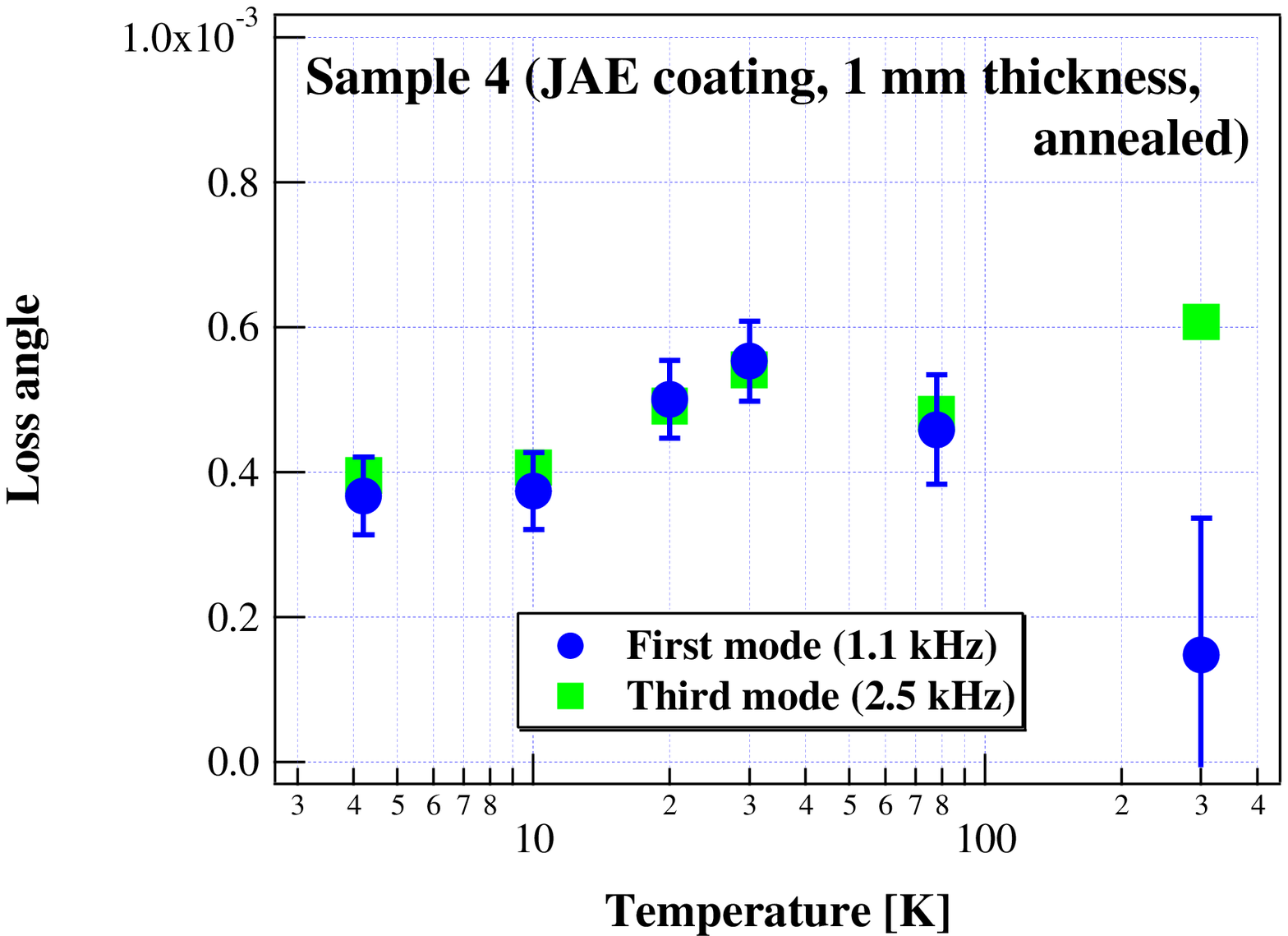}
\end{minipage}
\caption{\label{coating}Measured mechanical loss angles 
of the coating as a function of the temperature. 
The circles (blue in online) and squares (green in online) 
represent the coating loss angles evaluated 
from the Q-values of the first and third modes, respectively.}
\end{figure*}
The coating loss angles derived from the measured Q-values are shown in 
Fig. \ref{coating} \cite{error}. 
The circles (blue in online) and squares (green in online) 
represent the coating loss angles evaluated 
from the Q-values of the first and third modes, respectively. 
The four samples had similar loss angles. 
The loss did not strongly depend 
on the temperature and resonant frequency. 
These loss angles were $(4 \sim 6) \times 10^{-4}$.
(Some loss angles at 300 K had the large error bars.  
The upper limits of these error bars were the order of $10^{-4}$.)


\section{Discussion}


\subsection{Properties of the coating mechanical loss}

Figure \ref{coating} implies that the coating loss is almost independent of 
the temperature between 4 K and 300 K. At room temperature, 
the coating loss is dominated by that of Ta$_2$O$_5$ \cite{Penn}. 
Thus, it is expected that the loss of Ta$_2$O$_5$ is also
the main component of the coating loss, and is constant 
in the low-temperature region.  
The loss of the SiO$_2$ does not change very much, either. 
Research showed that 
the loss of the fused silica film deposited by 
the electron beam evaporation 
was independent of the temperature \cite{White}.    
The temperature dependence of  
these two kinds of thin SiO$_2$ is different from that of bulk silica. 
Many studies (for example, Refs. \cite{Fine, Strakna}) 
found that the Q-value of 
the bulk SiO$_2$ has a local minimum (about $10^{3}$) of 
around 30 K. 

Recently, the thermoelastic damping of the coating at room temperature 
was investigated theoretically
\cite{Braginsky-coating, Fejer}. 
Since the thermoelastic damping strongly depends on temperature 
in general, our result suggests that the contribution 
of thermoelastic damping is not a large part in the loss 
of the coating. 

The measured cold coating loss angles were not
affected by a change in the frequency between 0.52 kHz and 
2.5 kHz. 
This result suggests that the loss which depends on the frequency 
(for example, the thermoelastic damping) does not dominate the coating loss 
around this frequency region. 


The coating applies a stress on the substrate of the mirror. 
If the strain caused by this stress causes a loss, 
the coating loss on the thin disk in our experiment is different 
from that on thick mirrors. Fortunately, this scenario is 
rejected because the measured coating loss on a 0.5 mm thick disk 
was the same as that on a 1 mm thick disk.

It was reported that annealing improves the Q-values of 
fused silica \cite{Fraser,PennFS,Numata4}. 
There was no such effect on the coating in our samples. 
This result implies that the stress does not greatly change the loss 
because the annealing relaxes the stress produced 
during the coating process.

The loss of the NAOJ coating 
was about the same as that of the JAE coating. 
Since the optical loss of the NAOJ coating was larger than 
that of the JAE coating, the source of the optical loss 
does not have a large contribution on the mechanical loss.

\begin{figure}
\includegraphics[width=8.6cm]{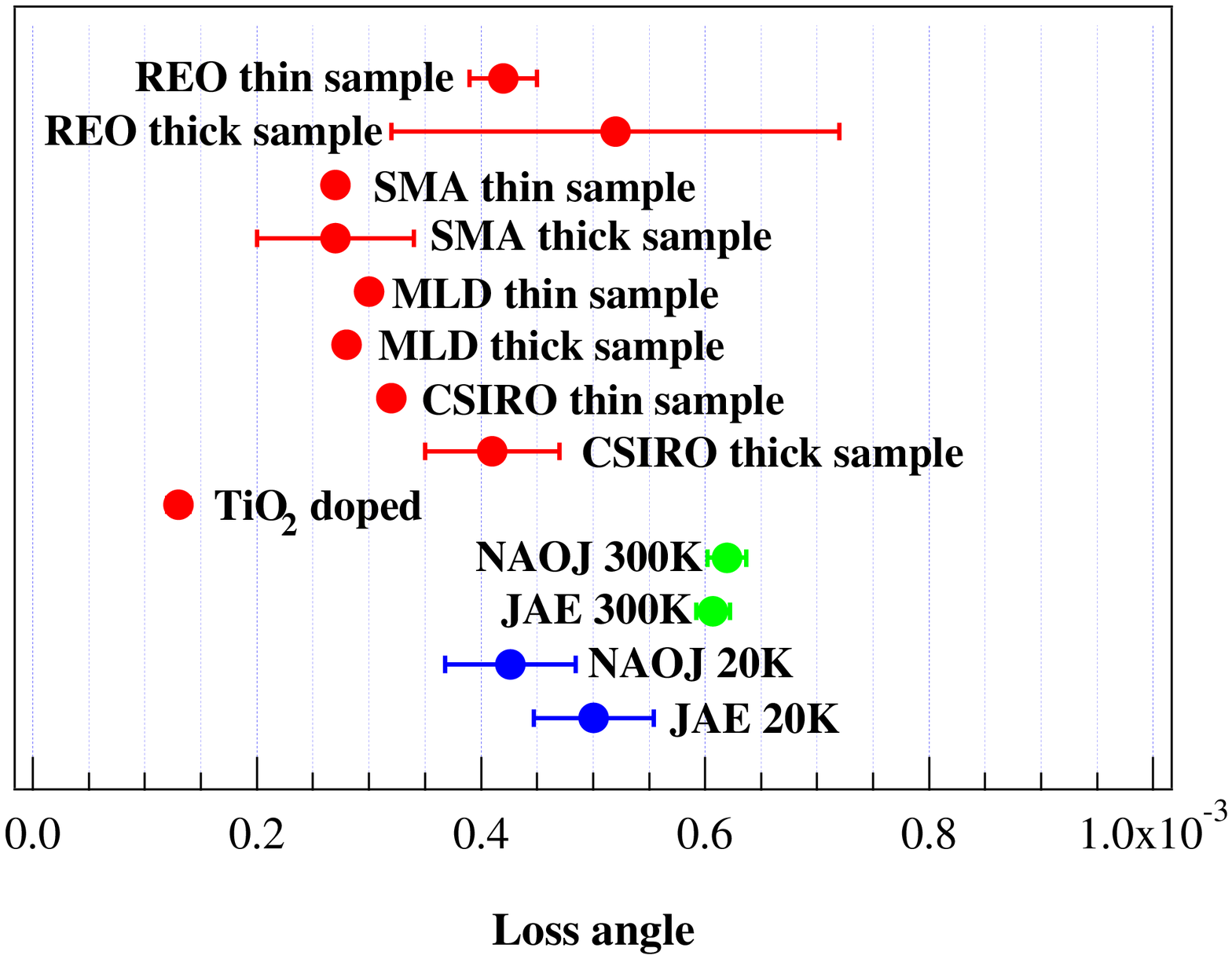}
\caption{\label{other}Summary of 
our measured coating loss with those of other groups (SiO$_2$/Ta$_2$O$_5$). 
The experiments of the other groups were at room temperature. 
Abbreviations show venders (REO: \cite{REO}, SMA: \cite{SMA}, 
MLD: \cite{MLD}, CSIRO: \cite{CSIROaddress}). 
The references are as follows: 
REO thin sample: \cite{Harry}, 
REO thick sample: \cite{Harry,Penn},
SMA thin sample, SMA thick sample, 
MLD thin sample, MLD thick sample: 
\cite{Penn} (The error bars of the SMA thin and both MLD samples 
are not shown in Ref. \cite{Penn}. 
According to Ref. \cite{Crooks-Amaldi}, 
the errors on the thin samples were $\pm 0.3 \times 10^{-4}$ at most. 
Probably, the error of the MLD thick sample 
was about the same as that of the SMA thick one 
because the substrate was the same.), 
CSIRO thin sample, 
CSIRO thick sample: \cite{CSIRO} (The error of 
the thin sample is not written in Ref. \cite{CSIRO}. However, 
this error was comparable to those of the SMA and MLD thin ones, 
$\pm 0.3 \times 10^{-4}$, probably.), 
TiO$_2$ doped: \cite{Cagnoli-Amaldi} 
(The error was $\pm 0.1 \times 10^{-4}$.),
NAOJ 300K, JAE 300K, NAOJ 20K, JAE 20 K: our results.}
\end{figure}
In Fig. \ref{other}, a summary of our results 
with those of the measurements 
by other groups (SiO$_2$/Ta$_2$O$_5$) is listed 
(their references are written in the figure caption). 
The experiments of the other groups were at room temperature. 
The losses at 300 K 
in the coatings prepared by the various companies were on the order of  
$10^{-4}$. The 
losses of the coating made by two laboratories, NAOJ and JAE, 
were almost the same in the low-temperature region. 
Thus, it is expected that 
the details of 
the coating manufacturing processes do not greatly affect 
the coating loss between 4 K and 300 K. 
This conclusion is in contrast with 
that of bulk fused silica: 
in Ref. \cite{Numata4}, 
the best Q-value of the room temperature silica is about fifty-times larger 
than the worst one.

\subsection{Coating thermal noise in interferometric 
gravitational wave detector}



Our experiment shows that the coating loss is almost independent 
of the temperature between 4 K and 300 K. 
This result and the formulae in Refs. \cite{Harry,Nakagawa} imply that 
the amplitude of the thermal noise caused by the coating loss 
is proportional to the square root of the temperature. 
The thermal noises at 20 K and 4 K are four and ten-times less 
than that at 300 K, respectively.
Since the loss of sapphire decreases at low temperature, 
cooling reduces the thermal noise of the coating more modestly 
than that of the sapphire substrate loss. 
However, the other methods used to suppress the coating thermal noise 
(reduction of the loss, other coating material, 
large scale laser beam, some ideas about coating thickness) 
are not more effective than the cooling, as follows. 

In spite of investigations concerning a 
reduction of the coating mechanical loss, 
a method used to drastically suppress 
the loss was not found as shown in Fig. \ref{other}. 
Even if TiO$_2$ is doped \cite{Cagnoli-Amaldi}, 
the thermal noise becomes about two-times smaller, 
because it is proportional to 
the square root of 
the loss angle \cite{Harry,Nakagawa}. 
In our measurement, annealing was not useful. 

Studies about other coating materials
are also in progress 
(for example, Nb$_2$O$_5$/SiO$_2$, 
Ta$_2$O$_5$/Al$_2$O$_3$, Al$_2$O$_3$/SiO$_2$ 
\cite{Sneddon-Aspen,Harry-LSC}). 
An obviously better material than SiO$_2$/Ta$_2$O$_5$ was not found. 

Adopting a larger beam is one of the methods 
because the coating thermal noise
is inversely proportional to the beam radius 
\cite{Levin,Harry,Nakagawa,Yamamoto2}. 
When the beam becomes larger, the mirror must also be greater, 
owing to the diffraction loss. 
Because of technological limits about the scale of the mirror, 
the maximum beam radius is about 6 cm. 
Since the typical beam radius of the current km-class interferometers 
is about 3 cm, 
the reduction factor of the thermal noise is about two. 
Recently, a new beam profile, a flat-topped beam \cite{Mexican1},  
was proposed to increase the beam scale without a large diffraction loss. 
The radius of this beam is 9 cm \cite{Mexican1}.
The thermal noise becomes three-times smaller 
\cite{NumataMexican}. 

In Ref. \cite{Khalili}, it is considered 
to decrease the number of the coating layers 
effectively (from about thirty layers to a few layers) 
by putting another mirror behind the end mirror. 
The reduction factor of this idea is about four, at most, 
because the thermal noise is proportional to 
the square root of the thickness of the coating \cite{Harry,Nakagawa}.
The thermal noise of the non-periodic coating 
(the optical thicknesses of the layers are different from that of each other) 
is 1.4-times less than that of the usual coating (the optical thicknesses of 
the layers are the same) \cite{Pinto}.



\subsection{Adequate mirror temperature 
for future gravitational wave detector projects}

The temperature dependence of the thermal noise of the mirrors 
was estimated in order to evaluate an adequate mirror temperature 
for future interferometric gravitational wave detector projects. 
The thermal noise of the mirrors is dominated 
by the contributions of the coating 
and the substrate losses \cite{Yamamoto2}.
The formula of the thermal noise of the coating derived 
in Ref. \cite{Nakagawa} was used.  
It was supposed that 
the coating loss angle is independent of the frequency and $4 \times 10^{-4}$. 
The substrate loss is the 
summation of the structure \cite{Bondu} 
and thermoelastic damping \cite{Braginsky,Cerdonio}. 
It was assumed that the substrate 
Q-values are $10^8$ \cite{Uchiyama2,Rowansapphire,Silica}. 
The length of the interferometer baselines and 
the beam radius at the mirrors were 3 km and 3 cm, respectively.
These were typical values. 

\begin{figure}
\includegraphics[width=8.6cm]{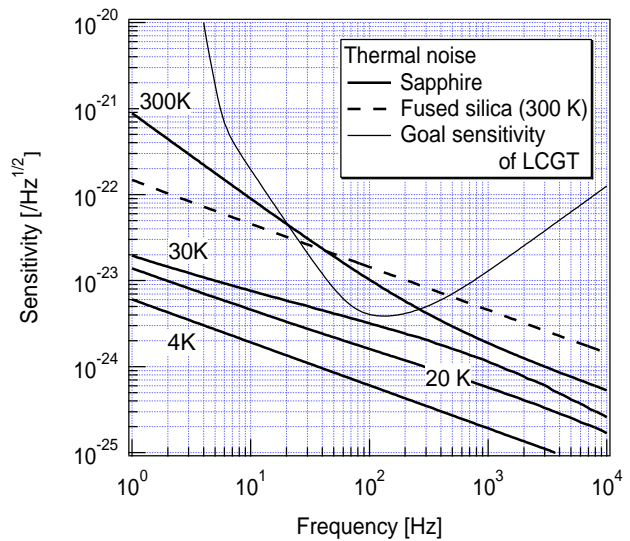}
\caption{\label{interferometer1}Thermal noise of 
the interferometric gravitational wave detector 
with 3 km baselines.  
The solid thick lines represent the thermal noises 
of the sapphire mirrors at 300 K, 30 K, 20 K, and 4 K. 
The dashed line is the thermal noise of mirrors made from  
fused silica, which is the material of the current interferometers 
\cite{LIGO,VIRGO,GEO,TAMA}, at 300 K as a reference.  
The thin line shows 
the goal sensitivity of the LCGT project \cite{LCGT} 
(the other future project, advanced LIGO \cite{LIGO II}, has 
a similar goal sensitivity).}
\end{figure}
The evaluated results are given in Fig. \ref{interferometer1}. 
The solid thick lines represent the thermal noises 
of the sapphire mirrors at 300 K, 30 K, 20 K, and 4 K. 
The dashed line is the thermal noise of mirrors made from  
fused silica, which is the material of the current interferometers 
\cite{LIGO,VIRGO,GEO,TAMA}, at 300 K as a reference.  
The thin line shows 
the goal sensitivity of the LCGT project \cite{LCGT} 
(the other future project, advanced LIGO \cite{LIGO II}, has 
a similar goal sensitivity).
The dominant loss component 
in each case is as follows. 
For the sapphire at room temperature, 
the thermoelastic damping and coating losses are the 
main component below and above 300 Hz, respectively.
For the sapphire at 30 K, 
the coating loss and thermoelastic damping 
are dominant below and above 40 Hz.
For the sapphire at 20 K, the coating loss is the main component, 
except at around 2 kHz. Only at around 2 kHz, 
is the contribution of the thermoelastic damping as large 
as that of the coating loss.
For the sapphire at 4 K and fused silica, the coating loss is dominant.

The thermal noise of the sapphire 
mirrors at 30 K is comparable to the LCGT sensitivity. The noise at 20 K 
is a few-times smaller. The LCGT mirrors must be below 20 K. 
The thermal noise at 20 K is ten-times less 
than that of the current detectors (fused silica, 300 K). 
If the LCGT mirrors are replaced by room-temperature fused-silica mirrors, 
the observable distance of the chirp wave from the $1.4 M_{\odot}$ 
neutron star 
binary coalescence becomes about 2.7-times shorter \cite{Binary range}.



\subsection{Thermal noise in laser frequency stabilization}

The thermal noise of the coating will be a serious problem 
in the frequency stabilization of the laser. 
Recent research \cite{Numata-freqstab} proved that the world-highest level 
of laser frequency stabilization using a rigid cavity 
is only three-times larger 
than the coating thermal noise. 
In the near future, the laser stabilization technique will achieve a 
fundamental physical limit, the coating thermal noise. 
According to our experiment, 
the coating thermal noise of the cavity at 4 K is 
ten-times smaller than that at 
room temperature. A cryogenic rigid cavity 
is one of the promising techniques 
to drastically improve the frequency stabilization. 

\section{Conclusion}

In order to effectively suppress the thermal noise 
of interferometric gravitational 
wave detectors, 
it was proposed to cool the mirrors. 
To evaluate the thermal noise of the cryogenic mirrors, 
the mechanical loss in the cooled mirrors must be investigated.  
However, there had been no report 
about the loss of the reflective coating (SiO$_2$/Ta$_2$O$_5$) 
at low temperature until our experiment. 
The coating loss measurement is also an interesting topic 
in material science. 

Our measured loss angles, $(4 \sim 6) \times 10^{-4}$,  
were almost constant between 
4 K and 300 K. 
Since the room-temperature coating loss 
is dominated by the loss of Ta$_2$O$_5$, 
it is expected that 
the loss of Ta$_2$O$_5$ is also 
the main component in the low-temperature region. 
The loss in the ion-beam sputtered thin Ta$_2$O$_5$ (and also SiO$_2$) layer 
is independent of the temperature. 
The measured coating mechanical loss 
was not affected by changes of the frequency, optical loss, 
or stress caused by the coating, and venders.    

Since the coating loss does not strongly depend on the temperature, 
the amplitude of 
the thermal noise of the coating is proportional to 
the square root of the temperature. 
As far as we know, 
there is no more effective method to suppress the 
coating thermal noise than cooling. 
The limit of laser frequency stabilization using a rigid cavity 
due to the coating thermal noise 
decreases by an order of magnitude when the rigid cavity is 
cooled from 300 K to 4 K. 
The amplitude of the total amount of the coating and substrate 
thermal noises of the sapphire mirrors at 20 K 
is sufficiently lower than the sensitivity of 
even future interferometric gravitational wave detector 
projects, for example, LCGT.  


\begin{acknowledgments}
This work was in part supported by the 2001st year Joint Research Project
(Soken/K00-3) of Sokendai (The Graduate University for Advanced
Studies) 
and a Grant-in-Aid for Scientific Research of the Ministry 
of Education, Culture, Sports, Science and Technology. 
The MOU exchanged by three directors, Institute for Cosmic Ray Research 
(the University of Tokyo), National Astronomical Observatory of Japan (NAOJ)
and High Energy Accelerator Research Organization, assisted this research. 
\end{acknowledgments}


\begin{thebibliography}{99}

\bibitem{LIGO}
A. Abramovici {\it et al.}, Science {\bf 256}, 325 (1992).

\bibitem{VIRGO}
C. Bradaschia {\it et al.}, Nucl. Instr. and Meth. in Phys. Res. A 
{\bf 289}, 518 (1990).

\bibitem{GEO}
B. Willke {\it et al.}, Class. Quantum Grav. {\bf 19}, 1377 (2002).

\bibitem{TAMA}
M. Ando {\it et al.}, Phys. Rev. Lett. {\bf 86}, 3950 (2001).

\bibitem{Numata-freqstab}
K. Numata, A. Kemery, and J. Camp, Phys. Rev. Lett. {\bf 93}, 250602 (2004).

\bibitem{Uchiyama1}
T. Uchiyama {\it et al.}, Phys. Lett. A {\bf 242}, 211 (1998).

\bibitem{LCGT}
T. Uchiyama {\it et al.}, Class. Quantum Grav. {\bf 21}, S1161 (2004).

\bibitem{EGO}
A. Giazotto and G. Cella, Class. Quantum Grav. {\bf 21}, S1183 (2004).

\bibitem{Uchiyama2}
T. Uchiyama {\it et al.}, Phys. Lett. A {\bf 261}, 5 (1999).

\bibitem{Levin}
Yu. Levin, Phys. Rev. D {\bf 57}, 659 (1998).

\bibitem{Harry}
G.M. Harry {\it et al.}, Class. Quantum Grav. {\bf 19}, 897 (2002).

\bibitem{Yamamoto-Amaldi}
K. Yamamoto, S. Otsuka, M. Ando, K. Kawabe, and K. Tsubono, 
Class. Quantum Grav. {\bf 19}, 1689 (2002).

\bibitem{Nakagawa}
N. Nakagawa, A.M. Gretarsson, E.K. Gustafson, and M.M. Fejer, 
Phys. Rev. D {\bf 65}, 102001 (2002).

\bibitem{Yamamoto2}
K. Yamamoto, M. Ando, K. Kawabe, and K. Tsubono, 
Phys. Lett. A {\bf 305}, 18 (2002).

\bibitem{Numata3}
K. Numata, M. Ando, K. Yamamoto, S. Otsuka, and K. Tsubono, 
Phys. Rev. Lett. {\bf 91}, 260602 (2003).


\bibitem{Yamamoto3}
K. Yamamoto, S. Otsuka, Y. Nanjo, M. Ando, and K. Tsubono,
Phys. Lett. A {\bf 321}, 79 (2004).


\bibitem{Black}
E.D. Black {\it et al.}, Phys. Lett. A {\bf 328}, 1 (2004).

\bibitem{Vu}
P.D. Vu, X. Liu, and R.O. Pohl, Phys. Rev. B {\bf 63}, 125421 (2001).

\bibitem{Shinkosya}
SHINKOSHA CO., LTD., 2-4-1 Kosugaya, Sakae-ku, Yokohama, 
Kanagawa 247-0007, Japan
(http://www.shinkosha.com/e/index.html).

\bibitem{Sato}
S. Sato {\it et al.}, Appl. Opt. {\bf 38}, 2880 (1999).


\bibitem{Numata}
K. Numata {\it et al.}, Phys. Lett. A {\bf 276}, 37 (2000).

\bibitem{transducer}
The shape of the electrode of 
the actuator and transducer was similar to that in Ref. \cite{Uchiyama2}. 
These electrodes were 40 mm $\times$ 40 mm. 
The distance between the disk and the electrodes 
was about 1 mm.

\bibitem{Support loss}
In this and preliminary experiments, 
the measured Q-values without the coating varied even though the same sample 
was measured. This was due to the contamination of the support system loss, 
probably, because this contamination strongly depends on the small difference 
between the center of the disk and the point grasped by the support system.
Since the displacement of the first mode near the disk center is larger 
than that of the third mode, the Q-values of the first mode were smaller.
According to our experiment, 
this support loss was not negligible when the measured Q-values were larger 
than the order of $10^{6}$. 
The stastical errors of the measured Q-values with the coating were small
because the coating loss 
was larger than the support loss (the Q-values were the order of $10^{5}$).  

\bibitem{Zener}
C. Zener, Phys. Rev. {\bf 52}, 230 (1937); {\bf 53}, 90 (1938).

\bibitem{Blair}
D.G. Blair and J. Ferreirinho, Phys. Rev. Lett. {\bf 49}, 375 (1982).

\bibitem{Crooks}
D.R.M. Crooks {\it et al.}, Class. Quantum Grav. {\bf 19}, 883 (2002).

\bibitem{Penn}
S.D. Penn {\it et al.}, Class. Quantum Grav. {\bf 20}, 2917 (2003).

\bibitem{Crooks-Amaldi}
D.R.M. Crooks {\it et al.}, Class. Quantum Grav. {\bf 21}, S1059 (2004).

\bibitem{Landau}
Equation (11.5) of L.D. Landau and E.M. Lifshitz, {\it Theory of Elasticity}
(Pergamon, New York, 1986).

\bibitem{formula}
This formula is valid when the coating is negligibly thin
compared with the substrate disk. 
The correction due to the anisotropic elasticity of the sapphire crystal and 
the difference between the Poisson ratio of the substrate 
and coating is negligible.

\bibitem{Sapieha}
J.E. Klemberg-Sapieha {\it et al.}, Appl. Opt. {\bf 43}, 2670 (2004). 

\bibitem{Martin}
P.J. Martin {\it et al.}, 
{\it Proceedings of Thin Films: Stresses and Mechanical
Properties IV, April 1993}, Mater. Res. Soc. Symp. Proc. 
{\bf 308}, 583 (1993). 

\bibitem{CSIRO}
G. Harry, in {\it LIGO Scientific Collaboration meeting, Hanford, Washington, 
USA, August 2004}, http://www.ligo.caltech.edu/docs/G/G040330-00/.

\bibitem{Young}
In Refs. \cite{Sapieha, Martin, CSIRO}, 
the Young's moduli of the thin SiO$_2$ and Ta$_2$O$_5$ were measured. 
The Young's modulus of the thin SiO$_2$ \cite{Sapieha} 
was the almost same as that of the bulk.
These values were used in the previous 
work about the coating for the gravitational wave detection 
\cite{Harry, Crooks, Penn, Crooks-Amaldi}. 
It was assumed that the Young's moduli 
are independent of the temperature 
because, in general, the temperature dependence is small
(the Young's modulus of the bulk SiO$_2$ was almost constant \cite{Fine}).



\bibitem{Fine}
M.E. Fine, H. Van Duyne, and N.T. Kenney, J. Appl. Phys. {\bf 25}, 402 (1954).

\bibitem{error}
The error bars in Fig. \ref{coating} show the statistical errors of 
the measured values.
In the low-temperature region, these were dominated by the deviation 
of the support loss \cite{Support loss} in most cases.
The error bars do not include the differences 
between the substrate loss in each sapphire disk. 
The effect of these differences was small because the evaluated coating loss 
angles of the four samples in Fig. \ref{coating} were almost the same.

\bibitem{White}
B.E. White, Jr. and R.O. Pohl, Phys. Rev. Lett. {\bf 75}, 4437 (1995).

\bibitem{Strakna}
R.E. Strakna, Phys. Rev. {\bf 123}, 2020 (1961).

\bibitem{Braginsky-coating}
V.B. Braginsky and S.P. Vyatchanin, Phys. Lett. A {\bf 312}, 244 (2003).

\bibitem{Fejer}
M.M. Fejer {\it et al.}, Phys. Rev. D {\bf 70}, 082003 (2004).

\bibitem{Fraser}
D.B. Fraser, J. Appl. Phys. {\bf 41}, 6 (1970).

\bibitem{PennFS}
S.D. Penn {\it et al.}, Rev. Sci. Instrum. {\bf 72}, 3670 (2001).

\bibitem{Numata4}
K. Numata {\it et al.}, Phys. Lett. A {\bf 327}, 263 (2004). 

\bibitem{REO}
Research Electro-Optics, Inc., 5505 Airport Blvd, Boulder, 
Colorado 80301, USA (http://www.reoinc.com/).

\bibitem{SMA}
{\it Service des Materiaux Avances/Virgo}, Lyon, France.

\bibitem{MLD}
MLD Technologies, 2672 Bayshore Parkway, Suite ${\sharp}$701, 
Mountain View, California 94043, USA (http://mldtech.com/).

\bibitem{CSIROaddress}
Commonwealth Scientific and Industrial Research Organisation, 
Telecommunications and Industrial Physics, Sydney, Australia
(http://www.csiro.au/).

\bibitem{Cagnoli-Amaldi}
G. Cagnoli {\it et al.}, {\it Proceedings of the 6th Edoardo Amaldi 
Conference on Gravitational Waves, Okinawa, Japan, June 2005}, 
J. Phys.: Conf. Ser. {\bf 32}, 386 (2006) 
[Institute of Physics, Bristol, U.K.].

\bibitem{Sneddon-Aspen}
P. Sneddon {\it et al.}, in {\it The 2003 Aspen Winter Conference 
on Gravitational Waves and their Detection, Aspen, 
Colorado, USA, February 2003}, 
http://www.ligo.caltech.edu/docs/G/G030195-00.pdf.

\bibitem{Harry-LSC}
G. Harry {\it et al.}, in {\it LIGO Scientific Collaboration meeting, 
Livingston, Louisiana, USA, March 2003}, 
http://www.ligo.caltech.edu/docs/G/G030036-00/.

\bibitem{Mexican1}
E. D'Ambrosio, Phys. Rev. D {\bf 67}, 102004 (2003).

\bibitem{NumataMexican}
K. Numata, in {\it The 2003 Aspen Winter Conference on Gravitational Waves 
and their Detection, Aspen, Colorado, USA, February 2003}, 
http://www.ligo.caltech.edu/docs/G/G030213-00.pdf.

\bibitem{Khalili}
F.Ya. Khalili, Phys. Lett. A {\bf 334}, 67 (2005).

\bibitem{Pinto}
J. Agresti {\it et al.}, in {\it LIGO Scientific Collaboration meeting, 
Hanford, Washington, USA, August 2005}, 
http://www.ligo.caltech.edu/docs/G/G050363-00/.



\bibitem{Bondu}
F. Bondu, P. Hello, and J.-Y. Vinet, Phys. Lett. A {\bf 246}, 227 (1998).

\bibitem{Braginsky}
V.B. Braginsky, M.L. Gorodetsky, and S.P. Vyatchanin, Phys. Lett. A 
{\bf 264}, 1 (1999).

\bibitem{Cerdonio}
M. Cerdonio, L. Conti, A. Heidmann, and M. Pinard, Phys. Rev. D {\bf 63}, 
082003 (2001).

\bibitem{Rowansapphire}
S. Rowan {\it et al.}, Phys. Lett. A {\bf 265}, 5 (2000).

\bibitem{Silica}
A. Ageev, B.C. Palmer, A. De Felice, S.D. Penn, and P.R. Saulson, 
Class. Quantum Grav. {\bf 21}, 3887 (2004).

\bibitem{LIGO II}
P. Fritschel, {\it Proceedings of the SPIE meeting 
Gravitational-Wave Detection (4856-39), Waikoloa, Hawaii, 2002},
edited by P. Saulson and M. Cruise 
(International Society for Optical Engineering, WA, 2002), p.282.

\bibitem{Binary range}
L.S. Finn and D.F. Chernoff, Phys. Rev. D {\bf 47}, 2198 (1993).

\end{thebibliography}

\end{document}